\documentstyle[twoside,fleqn,espcrc2,epsfig,psfig]{article}
\def\cc{\langle{\bar \psi} \psi \rangle}
\def\be{\begin{equation}}
\def\ee{\end{equation}}

\def\lsim{\raise0.3ex\hbox{$<$\kern-0.75em\raise-1.1ex\hbox{$\sim$}}}
\def\gsim{\raise0.3ex\hbox{$>$\kern-0.75em\raise-1.1ex\hbox{$\sim$}}}

\def\NP{{ Nucl.\ Phys.\ }}
\def\PL{{ Phys.\ Lett.\ }}
\def\PR{{ Phys.\ Rev.\ }}

\def\PRL{{ Phys.\ Rev.\ Lett.\ }}

\newcommand{\AmS}{{\protect\the\textfont2

  A\kern-.1667em\lower.5ex\hbox{M}\kern-.125emS}}

\title{Deconfinement Through Chiral Transition In 2 Flavour QCD
\thanks{we are grateful for the financial support by the German Science 
Ministry BMBF under contract No. 06BI902.} 
}

\author{S. Digal, E. Laermann, H. Satz
\\\vskip 6pt
Fakult{\"a}t f{\"u}r Physik, 
    Universit{\"a}t Bielefeld,\\
    D-33501 Bielefeld, Germany} 
       
\begin{document}

\begin{abstract}
We propose that in QCD with dynamical quarks, colour deconfinement
occurs when an external field induced by the chiral
condensate strongly aligns the Polyakov loop. This effect sets in at the
chiral symmetry restoration temperature $T_{\chi}$ and thus makes
deconfinement and chiral symmetry restoration coincide. The predicted
singular behavior of Polyakov loop susceptibilities at $T_{\chi}$ is
shown to be supported by finite temperature lattice calculations.
\end{abstract}

\maketitle

\section{INTRODUCTION}
Hot QCD matter undergoes two well-defined phase transitions at zero baryon 
chemical potential depending on the bare quark mass parameter. For 
($m_q=\infty$) there is the deconfinement transition associated with the
spontaneous breaking of the global center symmetry $Z_{N_c} \in SU(N_c)$ of
the pure $SU(N_c)$ gauge theory. $N_c$ is the number of colours of
quarks. The order parameter describing this transition is 
the expectation value of the Polyakov
loop $L(T)$. There exists a critical temperature $T_d$, with 
$L(T) = 0 ~~\forall~~T \leq T_d$ and $L(T) > 0~~\forall~~T>T_d$. This 
transition is similar to the magnetic transition in $Z_N$ spin theories with
the magnetization $m(T)$ being analogous to $L(T)$. The deconfinement
transition and the transition in a spin system have same order when the
number of colours $N_c$ and the number of components of spin $N$ are equal
\cite{SY}. The nature of the deconfinement transition changes with $N_c$, for 
$N_c=2$ the transition is second order and for $N_c\ge3$ it is first order.

On the other hand for $m_q=0$, there is the chiral transition at the
critical temperature $T_{\it x}$ 
associated with the restoration of chiral symmetry which is spontaneously 
broken at low temperatures. The order parameter describing this transition
is the chiral condensate $K \equiv \cc$. The nature of the chiral transition
depends on the number of quark flavours $N_f$. For $N_f=2$ the
transition is second order and first order for $N_f\ge 3$. In the case
of a second order chiral transition the critical behaviour of the chiral
condensate are described the critical exponents defined in the following.

\be
K(T,m_q=0) \sim (T_{\it x}-T)^{\beta},~~T \lsim T_{\it x} \label{2}
\ee
while at $T=T_{\it x}$, it vanishes as
\be
K(T_{\it x}=0,m_q) \sim m_q^{1/\delta} \label{3}
\ee
when $m_q \to 0$. The chiral quark mass susceptibility
\begin{eqnarray}
\it x_m^K(T,m_q) = \left( {\partial K \over \partial m_q}
\right)_T &\sim& ~ \langle  (\psi{\bar \psi})^2 \rangle - \langle
\psi{\bar \psi} \rangle^2~\nonumber \\
&\sim&~|T-T_{\it x}|^{-\gamma}
\label{4}
\end{eqnarray}
diverges at $T = T_{\it x}$ for $m_q \to 0$.
$\beta,~\delta,~\gamma,~...$
are the critical exponents.
It is believed that for $N_f=2$ the chiral
transition is in the universality class of $O(4)$ according to symmetry 
arguments \cite{PW}, however it is yet to be confirmed by numerical 
experiments \cite{Aoki}.

Both the symmetries of the Lagrangian, the global $Z_{N_c}$ symmetry and the
chiral symmetry, are explicitly broken for finite non-zero $m_q$. The role of 
$m_q$ for chiral transition is similar to that of an external magnetic field
in spin system, so that for large $m_q$, the temperature variation of the 
chiral condensate becomes completely smooth, like the
magnetization $m(T)$ in spin systems. 

It is not clear what role $m_q$ plays for the deconfinement transition. 
One might consider the inverse bare quark mass to play such a symmetry
breaking role for deconfinement, which would imply that for $m_q \to 0$,
the temperature variation of the Polyakov loop expectation value would
then become smooth \cite{HKS}. In lattice studies it is found, however, that
this is not the case \cite{KL}: even for $m_q \to 0$, the Polyakov loop
varies sharply with temperature and the corresponding susceptibility
($\langle L^2 \rangle - \langle L \rangle^2$) peaks sharply at the
chiral restoration temperature $T_{\it x}$, which is considerably lower 
than $T_d$. In somewhat loose terminology, one describes this situation by 
saying that in QCD for $m_q \to 0$, deconfinement and chiral symmetry 
restoration coincide.

To understand the underlying reason for this coincidence of the 
critical temperature of chiral and deconfinement transitions we propose
\cite{dgl}that the explicit breaking of the $Z_N$ symmetry is due to the
effective external field which is inverse of the constituent quark mass 
$m_Q$ \cite{Gavai}. It is believed that the constituent quark mass depends
on the chiral condensate $K$, however, the exact dependence is not yet known.  
If the non-Goldstone hadron masses are related to the chiral condensate, 
e.g., if the nucleon mass is given by \cite{Ioffe}
\be
M_n \sim 3~m_Q \sim K^{1/3}, \label{1}
\ee
then chiral symmetry restoration with $K \to 0$  gives rise to a
sudden large increase of the external field $H \sim 1/m_Q$, forcing the
Polyakov loop to take a large non-zero value. We want to argue that it is this 
effect which causes a sharp variation of the Polyakov loop $L(T)$ at 
$T_{\it x} < T_d$, with the temperature dependence of $L(T)$
governed by that of the chiral condensate $K(T)$, which is in general 
different from that obtained in pure gauge theory at $T_d$.

Although we believe that this interplay of chiral symmetry will
in general do not depend on the value of $N_f$ and $N_c$, for simplicity 
and to compare with lattice results, we consider $N_f=2$ and $N_c=3$
\cite{dgl}. As we have already mentioned in this case the chiral transition is
second order in the chiral limit, and deconfinement is first order in
the limit of infinite heavy quarks. 

As indicated above, we now take the $Z_N$ structure of QCD to be that
of $Z_N$ spin theory in an external field $H$, determined by the
constituent quark mass. From Eq.\ (\ref{1}), $H$ and hence also the
Polyakov loop thus become functions of $K$: the quark mass dependence of
$L$ enters through $K(T,m_q)$, with $L(T,K)$. In spin systems
with external field, the magnetization $m(T,H)$ is for $H\not=0$ an
analytic function of $T$ and $H$. We therefore assume that $L(T,K)$ is
for $\infty>K>0$ an analytic function of $T$ and $K$, with the quark
mass dependence of $L$ entering through $K(T,m_q)$.

\medskip

Once the effective external field for deconfinement depends on the
chiral condensate $K$, the critical behavior of the chiral condensate will
be reflected in the behavior of the Polyakov loop at $T=T_{\it x}$. For
a first order chiral transition with a jump in the chiral condensate 
across the chiral transition, there will be a discontinuity in the
effective external field giving rise to a discontinuity in the Polyakov
loop across the chiral transition point \cite{Rajan}-\cite{MTc}. To go
further one can write down for the variation of 
Polyakov loop $L$,

\be
dL = \left( {\partial L \over \partial T} \right)_K dT +
\left( {\partial L \over \partial K} \right)_T dK.
\label{5}
\ee

\noindent From this we can calculate the variation of the Polyakov loop
with respect to $m_q$ and temperature T as follows,

\be
\it x^L_m = \left( {\partial L \over \partial m_q} \right)_T =
\left( {\partial L \over \partial K} \right)_T
\left({\partial K \over \partial m_q} \right)_T
\label{6a}
\ee

and

\be
\it x^L_T = \left( {\partial L \over \partial T} \right)_{m_q} =
\left( {\partial L \over \partial T} \right)_K +
\left( {\partial L \over \partial K} \right)_T
\left( {\partial K \over \partial T} \right)_{m_q}.
\label{6b}
\ee
Both these response functions of the Polyakov loop diverge at 
$T\!=\!T_{\it x}$ in the chiral limit $m_q\!=\!0$, since
the chiral susceptibilities $\it x_m^K = (\partial K / \partial m_q)_T$ and
$\it x_T^K = (\partial K/\partial T)_{m_q}$ diverge in this limit.

In \cite{KL}, the chiral susceptibilities $\it x_T^K$ and $\it x_m^K$
were studied on an $8^3\times 4$ lattice for quark masses
$m_qa= 0.075,~0.0375$ and 0.02. Results for $\it x_T^K$ and $\it x_m^K$
are shown in Figs. 1 and 2, respectively, as functions of the effective 
temperature variable $T\equiv \kappa=6/g^2$, where $g$ denotes the coupling in
the QCD Lagrangian. The peak heights of both these response functions
increase with decreasing quark mass, which
indicates the divergence in the chiral limit $m_q \to 0$. In Figs.
3 and 4, we show the corresponding results for the
Polyakov loop susceptibilities. They also peak sharply, the 

\begin{figure}[htb]
\epsfig{file=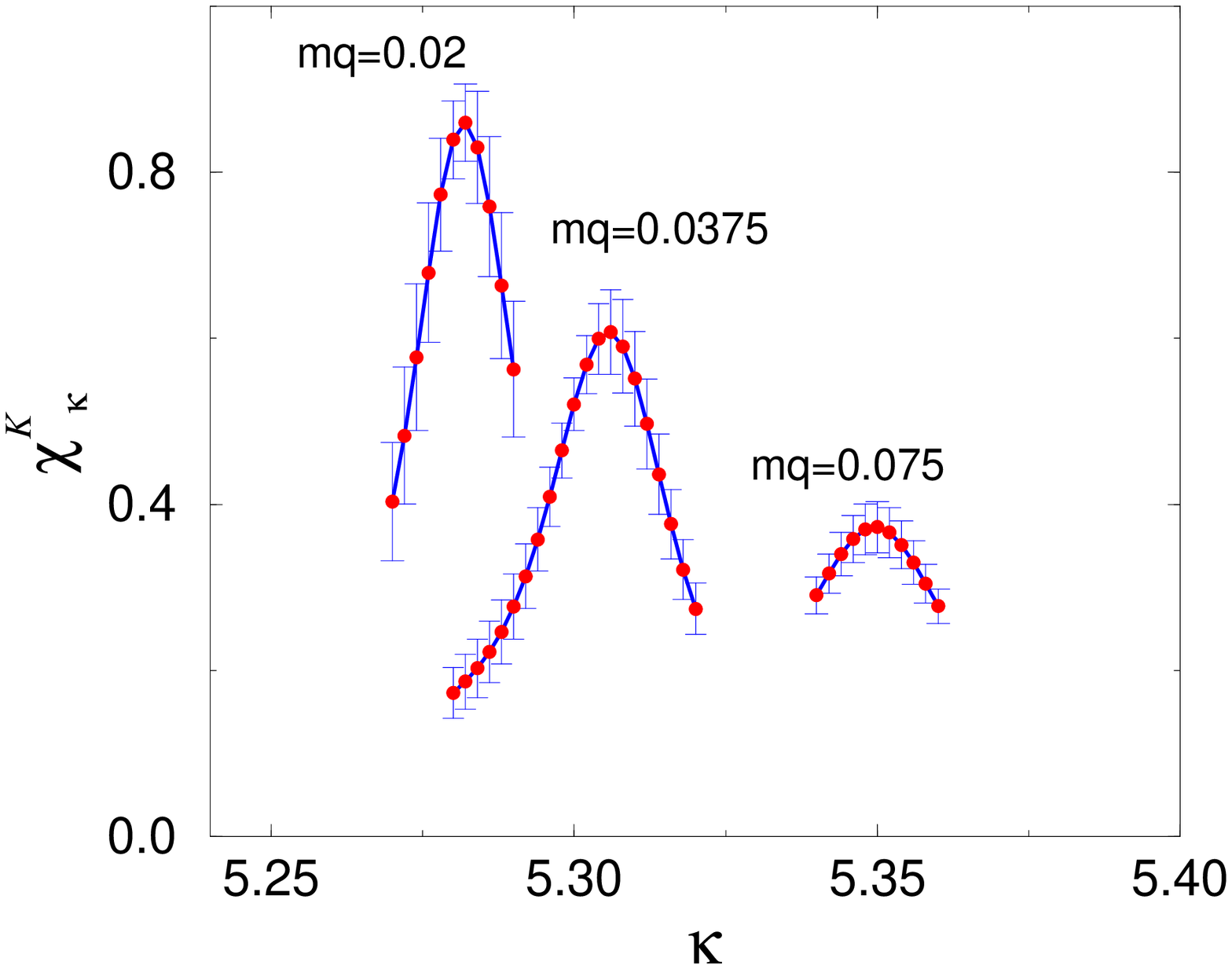,width=75mm}
\vspace{-12mm}
\caption{The chiral temperature
susceptibility $\it x^K_{\kappa}$ as function of the temperature variable
$\kappa=6/g^2$.}
\label{ek1}

\epsfig{file=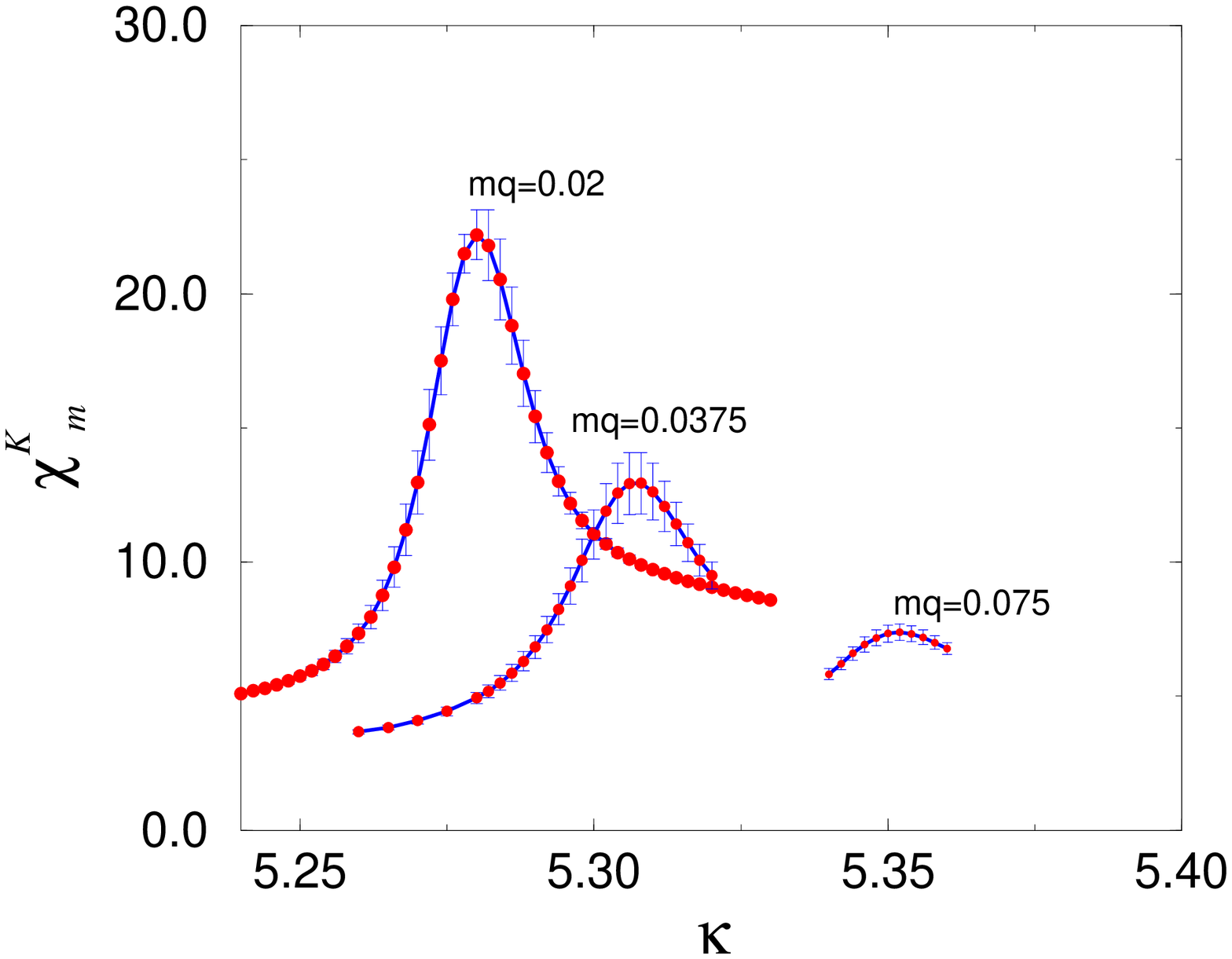,width=75mm}
\vspace{-12mm}
\caption{The chiral quark mass susceptibility $\it x^K_m$ as function of the
temperature variable $\kappa=6/g^2$.}
\label{do1}
\end{figure}


\begin{figure}[htb]
\epsfig{file=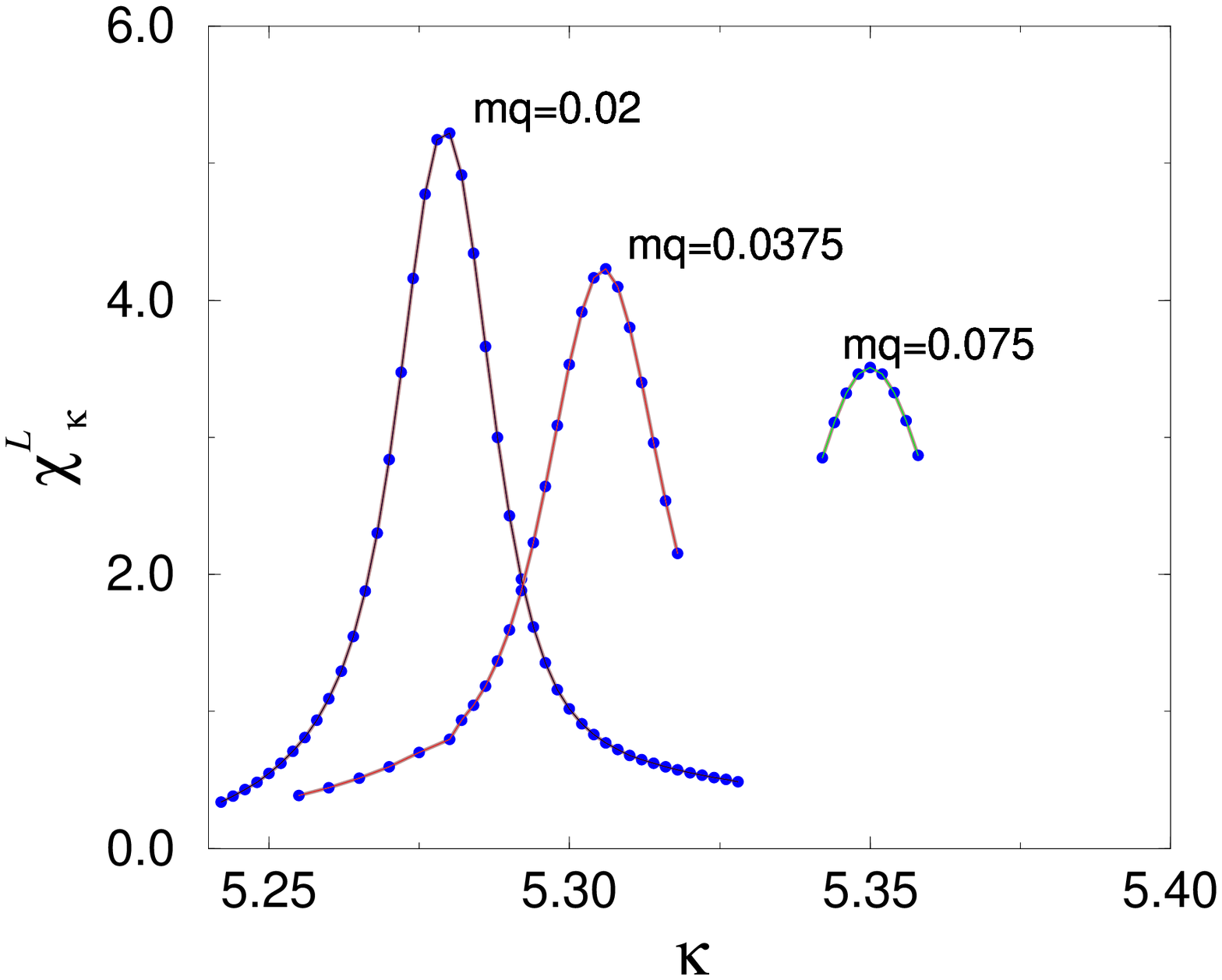,width=75mm}
\vspace{-12mm}
\caption{The Polyakov loop
temperature susceptibility $\it x^L_{\kappa}$ as
function of the temperature variable $\kappa=6/g^2$.}
\label{ek}

\epsfig{file=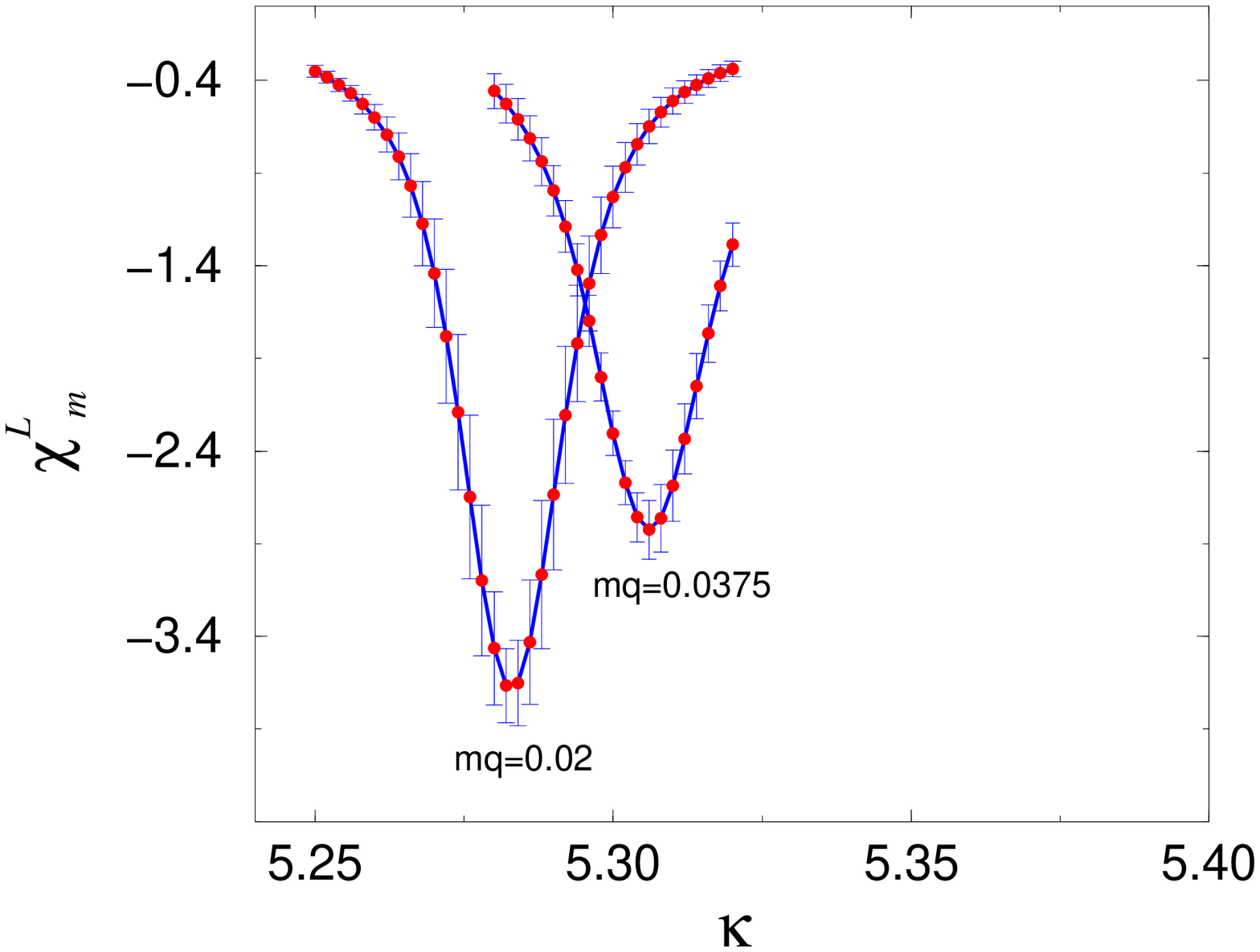,width=75mm}
\vspace{-12mm}
\caption{The Polyakov loop quark mass susceptibility $\it x^L_m$ as
function of the temperature variable $\kappa=6/g^2$.}
\label{do}
\end{figure}

\clearpage

\noindent peak positions coincide 
with those for the chiral susceptibilities, and here
as well the peak height increases with decreasing quark mass. The
observed behavior therefore provides clear support for the divergence
of the temperature and quark mass derivatives of the Polyakov loop at
$T_{\it x}$ in the chiral limit.

A direct comparison of the functional behavior, ideally of the relevant
critical exponents, for the two cases becomes difficult for two
reasons. The chiral susceptibilities do not at present lead to the
predicted $O(4)$ exponents; this may indicate that the quark masses are
still too large for pure critical behavior. The Polyakov loop
susceptibilities contain in addition unknown non-singular factors
$(\partial L/ \partial T)_K$ and $(\partial L/ \partial K)_T$, which
will modify the non-singular $m_q$-dependence relative to that of the
chiral susceptibilities. Lattice studies for smaller quark masses would
therefore be very helpful.

In summary, we have shown that chiral symmetry restoration, with the
resulting sudden change in an effective constituent quark mass, leads
to a suddenly increasing external field which aligns the Polyakov loops
at $T=T_{\it x} < T_d$
and thus produces a strong explicit breaking of the $Z_N$ symmetry of
the gauge field part of the Lagrangian. This effect makes chiral
symmetry restoration `coincide' with deconfinement. The resulting
predictions for diverging Polyakov loop susceptibilities are found to be
well supported by finite temperature lattice calculations for full QCD
with two flavours of light quarks.

\end{document}